\documentclass[aps,prl,preprint,amsmath,amssymb,showpacs,superscriptaddress]{revtex4}

\usepackage{graphicx}
\usepackage{dcolumn}
\usepackage{bm}
\usepackage{longtable}
\usepackage{color}
\usepackage{CJK}

\begin{document}

\title{Antimagnetic Rotation Band in Nuclei: A Microscopic Description}

\author{P. W. Zhao}%
\affiliation{State Key Laboratory of Nuclear Physics and Technology, School of Physics, Peking University, Beijing 100871, China}
\author{J. Peng}%
\affiliation{Department of Physics, Beijing Normal University, Beijing 100875, China}
\author{H. Z. Liang}%
\affiliation{State Key Laboratory of Nuclear Physics and Technology, School of Physics, Peking University, Beijing 100871, China}
\author{P. Ring}%
\affiliation{State Key Laboratory of Nuclear Physics and Technology, School of Physics, Peking University, Beijing 100871, China}
\affiliation{Physik Department, Technische Universit\"at M\"unchen, D-85747 Garching, Germany}
\author{J. Meng}%
\affiliation{State Key Laboratory of Nuclear Physics and Technology, School of Physics, Peking University, Beijing 100871, China}
\affiliation{School of Physics and Nuclear Energy Engineering, Beihang University, Beijing 100191, China}
\affiliation{Department of Physics, University of Stellenbosch, Stellenbosch, South Africa}

\date{\today}
\begin{abstract}
Covariant density functional theory and the tilted axis cranking method are used to investigate antimagnetic rotation (AMR) in nuclei for the first time in a fully self-consistent and microscopic way. The experimental spectrum as well as the $B(E2)$ values of the recently observed AMR band in $^{105}\rm Cd$ are reproduced very well. This gives a further strong hint that AMR is realized in specific bands in nuclei.
\end{abstract}

\pacs{21.60.Jz, 21.10.Re, 23.20.-g, 27.60.+j}

\maketitle

Similar to rotational bands observed in molecules, many nuclei have energy spectra with a pronounced rotational character. The most common rotational bands in nuclei are built on states with a substantial quadrupole deformation and show strong electric quadrupole ($E2$) transitions between the rotational states. Such bands are usually well interpreted as a coherent collective rotation of many nucleons around an axis perpendicular to the symmetry axis of the deformed density distribution~\cite{Bohr1975B}.

Over the past decades, the experimental discovery of regular rotational-like bands in nearly spherical lead isotopes and other groups of nuclei has opened a new era in high-spin physics (for reviews see Refs.~\cite{Hubel2005Prog.Part.Nucl.Phys.1,Frauendorf2001Rev.Mod.Phys.463,Clark2000Annu.Rev.Nucl.Part.Sci.1}).
Unlike the normal deformed rotational bands, these rotational-like bands have strong magnetic dipole ($M1$) and very weak $E2$ transitions. The orientation of these rotors is not specified by the deformation of the overall density but rather by the current distribution induced by specific nucleons moving in high-$j$ orbitals.

The explanation of such bands in terms of the ``shears mechanism'' was firstly given in Ref.~\cite{Frauendorf1993Nucl.Phys.A259}. The angular momentum vectors of the high-$j$ proton and neutron form two blades of a pair of shears and are almost perpendicular to each other at the bandhead. Along the bands, energy and angular momentum are increased by closing the blades of the shears, i.e., by alignment of the proton and neutron angular momenta. Consequently, the rotational bands are formed in spite of the fact that shapes of these nuclei stay nearly spherical. Clear evidence for this new rotation mode has firstly been provided through lifetime measurements for four $M1$ bands in $^{198,199}\rm Pb$~\cite{Clark1997Phys.Rev.Lett.1868}. In order to distinguish this kind of rotation from the usual collective rotation in well-deformed nuclei (called electric rotation), the name ``magnetic rotation'' (MR)~\cite{Frauendorf199452} was introduced, which alludes to the fact that the magnetic moment is the order parameter inducing a violation of rotational symmetry and thus causing rotational-like structures in the spectrum. This forms an analogy to a ferromagnet, where the total magnetic moment, the sum of the atomic dipole moments, is the order parameter.

\begin{figure}[htbp]
\includegraphics[width=6cm]{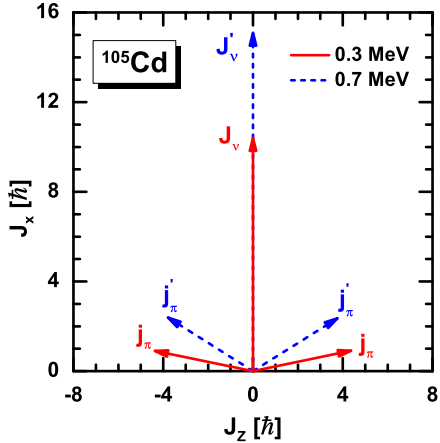}
\caption{(color online) Angular momentum vectors of neutrons $\bm{J}_\nu$ and the two $g_{9/2}$ proton holes $\bm{j}_\pi$ at both the bandhead ($\hbar\Omega=0.3$~MeV)
and the maximum rotational frequency.}
\label{fig1}
\end{figure}

In an antiferromagnet, on the other hand, one-half of the atomic dipole moments are aligned on one sublattice and the other half are aligned in the opposite direction on the second sublattice. Although there is no net magnetic moment in an antiferromagnet, the state is ordered; i.e., it breaks isotropy like a ferromagnet. In analogy with an antiferromagnet,
a similar phenomenon has been predicted in nuclei for ``antimagnetic rotation'' (AMR)~\cite{Frauendorf2001Rev.Mod.Phys.463}: in specific nearly spherical nuclei, subsystems of valence protons are aligned back to back in opposite direction and nearly perpendicular to the orientation of the total spin of the valence neutrons (see Fig.~\ref{fig1}). This arrangement also breaks rotational symmetry in these nearly spherical systems and causes excitations with rotational character on top of this bandhead. Along this band, energy and angular momentum are increased by a simultaneous closing of the two proton blades toward the neutron angular momentum vector. Consequently, a new kind of rotational band is formed showing some analogy with an antiferromagnet.

AMR is expected to be observed in the same regions as magnetic rotation~\cite{Frauendorf2001Rev.Mod.Phys.463}. However, it differs from magnetic rotation in two important issues. First, there are no $M1$ transitions since the transverse magnetic moments of the two subsystems are antialigned and cancel each other. Second, as the antimagnetic rotor is symmetric with respect to a rotation by $\pi$ about the angular momentum axis, the energy levels in the AMR band differ in spin by $2\hbar$ and are connected by weak $E2$ transitions reflecting the nearly spherical core. Moreover, the phenomenon of AMR is characterized by a decrease of the $B(E2)$ values with increasing spin, which can be demonstrated by measuring lifetimes. Since AMR was proposed~\cite{Frauendorf2001Rev.Mod.Phys.463}, it has attracted more and more interest both on the experimental and on the theoretical side. To date, firm  experimental evidence of AMR has been reported in $^{105,106,108}\rm Cd$~\cite{Simons2003Phys.Rev.Lett.162501,Simons2005Phys.Rev.C24318,Datta2005Phys.Rev.C41305,Choudhury2010Phys.Rev.C61308}, and the occurrence of this phenomenon in some other nuclei such as $^{109}\rm Cd$~\cite{Chiara2000Phys.Rev.C34318},
$^{100}\rm Pd$~\cite{Zhu2001Phys.Rev.C41302} has not yet been fully confirmed.

On the theoretical side, so far, AMR has been studied only by simple phenomenological models~\cite{Clark2000Annu.Rev.Nucl.Part.Sci.1}, in the framework of the pairing plus quadrupole model~\cite{Chiara2000Phys.Rev.C34318,Frauendorf2001Rev.Mod.Phys.463} and the microscopic-macroscopic model~\cite{Zhu2001Phys.Rev.C41302,Simons2003Phys.Rev.Lett.162501,Simons2005Phys.Rev.C24318,Datta2005Phys.Rev.C41305}. It has been shown that these models describe MR and AMR bands observed in the same nuclei~\cite{Chiara2000Phys.Rev.C34318,Frauendorf2001Rev.Mod.Phys.463}. However, polarization effects, which are expected to influence strongly the quadrupole moments and $B(E2)$ values, are either neglected completely or taken into account only partially by minimizing the rotating energy surface with respect to a few deformation parameters. The nuclear currents, which are the origin of symmetry violation in nuclei with AMR, are not treated in a self-consistent way in any of these models. It is evident that a full understanding requires self-consistent microscopic investigations including all degrees of freedom and based on reliable theories without additional parameters. Such calculations are not simple, but they are nowadays feasible in the framework of density functional theories (DFTs).

Such theories are used with great success in all quantum mechanical many-body systems, in particular, in Coulombic systems.
In nuclear physics with spin and isospin degrees
of freedom, the situation is much more complicated due to the strong nucleon-nucleon and three-body forces.
Covariant density functionals exploit basic properties of QCD at low energies, in particular, symmetries and the separation of scales~\cite{LNP.641}.
They provide a consistent treatment of the spin degrees of freedom, they include the complicated interplay between the large Lorentz scalar and vector self-energies induced on the QCD level by the in medium changes of the scalar and vector quark condensates~\cite{CFG.92} and they include the nuclear currents induced by of the spatial parts of the vector self-energies, which play an essential role in rotating nuclei. Of course, at present, all attempts to derive these functionals directly from the bare forces do not reach the required accuracy, but by fine-tuning a few phenomenological parameters to properties
of nuclear matter and finite nuclei, universal density functionals have been derived in recent years which provide an excellent description of ground states and excited states all over the periodic table with a high predictive power.


The tilted axis cranking (TAC) model based on relativistic~\cite{Peng2008Phys.Rev.C24313,Zhao2011Phys.Lett.B181,Madokoro2000Phys.Rev.C61301} and nonrelativistic~\cite{Olbratowski2004Phys.Rev.Lett.52501} density functionals has been applied in the past successfully for
magnetic rotation and chirality. This Letter presents the first fully self-consistent microscopic investigation of AMR in the framework of the TAC method based on covariant DFT.
This theory starts from a Lagrangian and the corresponding Kohn-Sham equations have the form of a Dirac equation with effective fields $S(\bm{r})$ and $V^\mu(\bm{r})$ derived from this Lagrangian. In the TAC model these fields are slightly deformed and the calculations are carried out in the intrinsic frame rotating with a constant angular velocity vector $\bm{\Omega}$ pointing in a direction which is not parallel to one of the principal axes of the density distribution:
\begin{equation}\label{Diracequation}
   [\bm{\alpha}\cdot(\bm{p}-\bm{V})+\beta(m+S)
    +V-\bm{\Omega}\cdot\hat{\bm{J}}]\psi_k=\epsilon_k\psi_k.
 \end{equation}
Here $\hat{\bm{J}}=\hat{\bm{L}}+\frac{1}{2}\hat{\bm{\Sigma}}$ is the total angular momentum of the nucleon spinors, and the fields $S$ and $V^\mu$ are connected in a self-consistent way to the densities and current distributions, for details see Refs.~\cite{Peng2008Phys.Rev.C24313,Zhao2011Phys.Lett.B181}. The iterative solution of these equations yields single-particle energies, expectation values of three components $\langle J_i\rangle$ of the angular momentum, energy,  quadrupole moments, $B(E2)$ transition probabilities, etc. The magnitude of the angular velocity $\Omega$ is connected to the total angular momentum quantum number $I$ by the semiclassical relation $\langle\hat{\bm{J}}\rangle\cdot\langle\hat{\bm{J}}\rangle=I(I+1)$.

We use the point coupling Lagrangian PC-PK1~\cite{Zhao2010Phys.Rev.C54319} and investigate the newly observed AMR band in the odd-A nucleus $^{105}\rm Cd$.
The calculations are free of additional parameters, pairing correlations are neglected.


In this work, we allow only rotations around an axis in the ($x,z$)-plane (2D cranking). Equation~(\ref{Diracequation}) is solved in a 3D Cartesian harmonic oscillator basis~\cite{Koepf1989Nucl.Phys.A493,Peng2008Phys.Rev.C24313} with $N=$10 major shells. In the experiment of $^{105}\rm Cd$ in Ref.~\cite{Choudhury2010Phys.Rev.C61308}, a negative-parity band was reported in forming the yrast line above $I = 23/2\hbar$. It was assumed that it corresponds to three aligned neutrons with the configuration $\nu[h_{11/2}(g_{7/2})^2]$ coupled to a pair of $g_{9/2}$ proton holes.


In the present calculations we consider, as in Ref.~\cite{Choudhury2010Phys.Rev.C61308}, the band where the odd neutron occupies the lowest level in the $h_{11/2}$ shell. The remaining nucleons are treated self-consistently by filling the orbitals according to their energy from the bottom of the well. This leads automatically at the bandhead to a configuration shown in Fig.~\ref{fig1}. In order to demonstrate the two ``shearslike'' mechanism in $^{105}\rm Cd$, we show both at the bandhead and at the maximum rotational frequency the angular momentum vectors of the two $g_{9/2}$ proton-holes $\bm{j}_\pi$ and of the neutrons $\bm{J}_\nu=\sum_{n}\bm{j}^{(n)}_\nu$
where $n$ runs over all the occupied neutron levels. At the bandhead, the two proton angular momentum vectors $\bm{j}_\pi$ are pointing opposite to each other and are nearly perpendicular to the vector $\bm{J}_\nu$. They form the blades of the two shears. With increasing $\Omega$ the gradual alignment of the vectors $\bm{j}_\pi$ of the two $g_{9/2}$ proton holes toward the vector $\bm{J}_\nu$ generates angular momentum while the direction of the total angular momentum stays unchanged. This leads to the closing of the two shears. The two shearslike mechanism can thus be clearly seen, and it is consistent with the previous works~\cite{Frauendorf2001Rev.Mod.Phys.463,Simons2003Phys.Rev.Lett.162501}.

\begin{figure}[htbp]
\includegraphics[width=8cm]{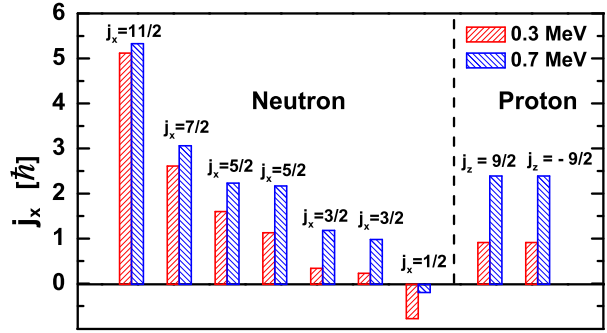}
\caption{(color online) Alignment of the valence neutrons (left side) and proton holes (right side) at both the bandhead ($\hbar\Omega=0.3$~MeV)
and the maximum rotational frequency.}
\label{fig2}
\end{figure}

In a microscopic calculation we have no inert core, all the energy and angular momentum comes from the particles. We show in Fig.~\ref{fig2} the contributions of the valence neutrons and proton holes to the angular momentum $J_x$ at both the bandhead and the maximum rotational frequency. It is found that the contributions come mainly from high-$j$ orbitals, i.e., from $g_{9/2}$ proton holes as well as from $h_{11/2}$ and $g_{7/2}$ neutrons. In order to provide a simple picture which can be compared with the core angular momentum given in
Ref.~\cite{Frauendorf2001Rev.Mod.Phys.463}, one can estimate the ``corelike'' angular momentum in the present framework by excluding the contributions of three valence neutrons, shown in the left three columns in Fig.2, from the total neutron angular momentum. It is found that the ``core'' contributes about 3 $\hbar$ when the frequency $\Omega$ increases from the bandhead to the maximum value.

For the protons, only the two holes in the $g_{9/2}$ shell contribute. As shown in Fig.~\ref{fig1}, they cancel each other in the $z$ direction giving non-negligible contributions to the angular momentum along $x$ axis even at the bandhead. With growing frequency, the proton angular momentum in $x$ direction increases because of the alignment of the two proton hole blades. For the neutrons, on the other hand, we have only contributions above the $N=50$ shell. One neutron sits in the $h_{11/2}$ orbit and the other six are, because of considerable mixing, distributed over the $g_{7/2}$ and $d_{5/2}$ orbitals.  As $\Omega$ grows, the contributions of the aligned orbitals with $j_x=11/2$ and $9/2$ do not change much and the increase in angular momentum is generated mostly by the alignment of orbitals with low $j_x$ values. This microscopic calculation shows that the phenomenological interpretation given in Ref.~\cite{Choudhury2010Phys.Rev.C61308} is only partially justified: we clearly have two proton holes in the $g_{9/2}$ and one neutron particle in the $h_{11/2}$ orbit, but, due to the mixing of orbits with lower $j$ values the other neutrons are distributed over several subshells above the $N=50$ core and the increasing angular momentum results from the alignment of the proton holes and the mixing within the neutron orbitals. Because of this strong mixing between the neutrons, a core needed for the phenomenological model in Ref.~\cite{Clark2000Annu.Rev.Nucl.Part.Sci.1} cannot really be defined.

\begin{figure}[htbp]
\includegraphics[width=7cm]{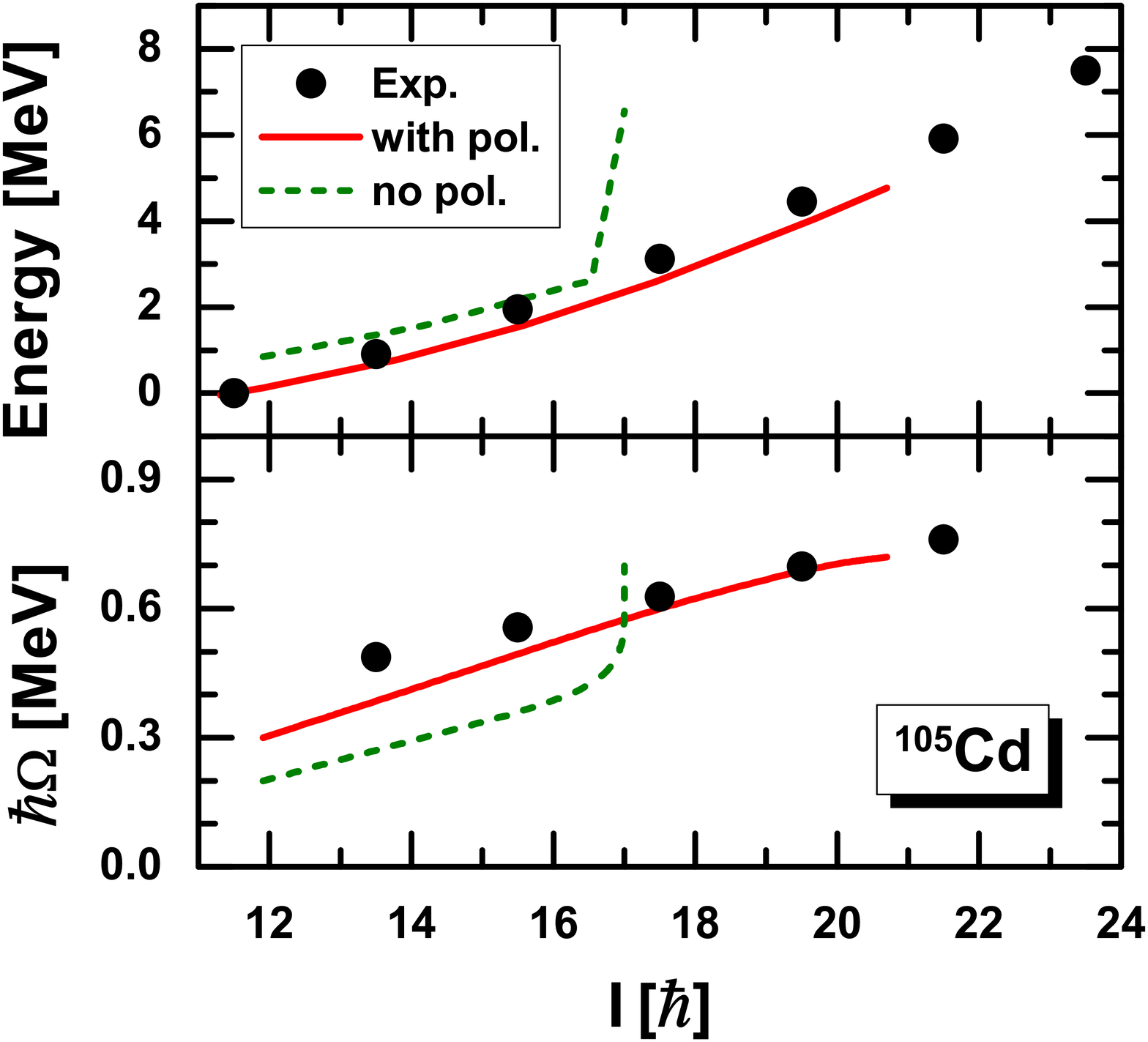}
\caption{(color online) Energy (upper panel) and rotational frequency (lower panel) as functions of the total angular momentum. The fully self-consistent solution (solid lines) and that neglecting polarization (dashed lines) are compared with the data~\cite{Choudhury2010Phys.Rev.C61308} (solid dots). The energy at $I = 23/2\hbar$ is taken as reference in the upper panel.}
\label{fig3}
\end{figure}

In Fig.~\ref{fig3} the calculated energy and the rotational frequency (solid lines) are compared with data~\cite{Choudhury2010Phys.Rev.C61308}. In the upper panel it can be clearly seen that, apart from the bandhead, the experimental energies are reproduced excellently by the present self-consistent calculations. In the lower panel it is found that the calculated total angular momenta agree well with the data and increase almost linearly with increasing frequency. This indicates that the  moment of inertia is nearly constant and well reproduced by the present calculations.

\begin{figure}[htbp]
\includegraphics[width=7cm]{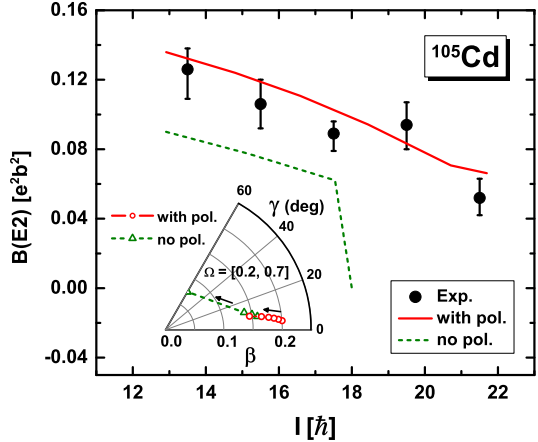}
\caption{(color online) $B(E2)$ values as a function of the angular momentum. Solutions with (solid line) and without (dashed line) polarization are compared with data~\cite{Choudhury2010Phys.Rev.C61308} (solid dots). Inset: Deformations $\beta$ and $\gamma$ driven by the increasing rotational frequency whose direction is indicated by arrows.}
\label{fig4}
\end{figure}

AMR is characterized by weak $E2$ transitions decreasing with increasing spin. In Fig.~\ref{fig4} we compare calculated $B(E2)$ values (solid lines) with the available data~\cite{Choudhury2010Phys.Rev.C61308}. It is found that the resulting $B(E2)$ values are very small ($< 0.14~e^2b^2$) and in very good agreement with the data. Furthermore, the fact that the $B(E2)$ values decrease with increasing spin is in agreement with the interpretation by the shearslike mechanism.

The decrease of the $B(E2)$ values can be understood by the changes in the nuclear deformation. As shown in the inset of Fig.~\ref{fig4}, with increasing frequency, the nucleus undergoes a rapid decrease of $\beta$ deformation from $0.2$ to $0.14$ at a small and near-constant triaxiality ($\gamma\leq 9^\circ$). As usual, it is found that the deformation of the charge distribution, responsible for the $B(E2)$ values, changes in a similar manner. Therefore, one can conclude that the alignment of the proton and neutron angular momenta, i.e., the two shearslike mechanism, is accompanied by a transition from prolate towards nearly spherical shape.

This is, however, not the full story. In order to investigate the importance of the polarization effects induced by the two proton holes, which is taken into account fully in the present work, an additional calculation without polarization has been carried out. For this purpose, we first calculated at each frequency the $^{107}\rm Sn$ core, where the two proton holes are filled. This results in a filled nearly spherical $g_{9/2}$ shell. In the next step we neglected self-consistency (dashed lines in Figs.~\ref{fig3} and \ref{fig4}) and calculated the band in $^{105}\rm Cd$ removing two protons from the $g_{9/2}$ shell using, at each frequency, the corresponding nearly spherical potentials $S$ and $V^\mu$ obtained in the calculations of the $^{107}\rm Sn$ core. As we can see in the upper panel of Fig.~\ref{fig3}, the energy is reduced only slightly by polarization in the lower part of the spectrum. At the same time the relation between angular velocity and angular momentum is considerably changed in the lower part of Fig.~\ref{fig3}. Without polarization we need a much smaller frequency $\Omega$ to reach the same angular momentum as with polarization. In addition, without polarization, there is a maximal angular momentum of roughly 17 $\hbar$. Higher values cannot be reached even at rather high frequencies.

This behavior can be well understood from the evolution of the deformation parameters shown in the inset of Fig.~\ref{fig4}. Without polarization we use at each frequency the potentials of the core-nucleus $^{107}\rm Sn$, where the deformation is relatively small. Angular momentum can only be produced by alignment of neutron particles along the rotational axis leading at $\hbar\Omega=0.5$ MeV to an oblate shape with a rotation around the symmetry axis. Removing two protons would lead, if polarization is taken into account, to a larger prolate configuration with lower energy and hindering alignment. Therefore, in the lower part of the band, it is easier to produce angular momentum without polarization, where the deformation is small. On the other side the oblate deformation keeps the high-$j$ proton holes in the $j_x=\pm 1/2$ orbitals of the $g_{9/2}$ shell pairwise occupied and hinders their alignment. Above $\hbar\Omega=0.5$ MeV we reach the maximum angular momentum of the neutron configuration. With polarization, because of the prolate deformation, we have much more mixing and therefore can reach larger angular momentum by aligning the protons. This can also be seen in Fig.~\ref{fig4} that the $E2$ transitions do survive with polarization when $\hbar\Omega\ge0.5$ MeV. Therefore, it is of importance to emphasize that polarization effects play a very important role in the self-consistent microscopic description of AMR bands.


In summary, for the first time, a band with antimagnetic rotation has been investigated in a fully self-consistent microscopic way by the TAC method based on covariant DFT. Without any additional parameters, the experimental energy spectrum and the $B(E2)$ values for the newly observed AMR band in $^{105}\rm Cd$ are reproduced very well when polarization effects are taken into account. The appearance of the two shearslike mechanism in AMR can be clearly seen in the present calculations. It is found that polarization effects play a very important role. Through more mixing they enhance alignment and they influence the deformation and $E2$ transitions.


This work was supported by the Major State 973 Program 2007CB815000, the NSFC (10975007, 10975008, 11005069, 11175002, 11105006), CPSF (20100480149), the Fundamental Research Funds for the Central University, and the DFG Cluster of Excellence ``Origin and Structure of the Universe.''


\begin{thebibliography}{21}
\expandafter\ifx\csname natexlab\endcsname\relax\def\natexlab#1{#1}\fi
\expandafter\ifx\csname bibnamefont\endcsname\relax
  \def\bibnamefont#1{#1}\fi
\expandafter\ifx\csname bibfnamefont\endcsname\relax
  \def\bibfnamefont#1{#1}\fi
\expandafter\ifx\csname citenamefont\endcsname\relax
  \def\citenamefont#1{#1}\fi
\expandafter\ifx\csname url\endcsname\relax
  \def\url#1{\texttt{#1}}\fi
\expandafter\ifx\csname urlprefix\endcsname\relax\def\urlprefix{URL }\fi
\providecommand{\bibinfo}[2]{#2}
\providecommand{\eprint}[2][]{\url{#2}}

\bibitem[{\citenamefont{Bohr and Mottelson}(1975)}]{Bohr1975B}
\bibinfo{author}{\bibfnamefont{A.}~\bibnamefont{Bohr}} \bibnamefont{and}
  \bibinfo{author}{\bibfnamefont{B.~R.} \bibnamefont{Mottelson}},
  \emph{\bibinfo{title}{Nuclear Structure}}, vol.~\bibinfo{volume}{II}
  (\bibinfo{publisher}{Benjamin, New York}, \bibinfo{year}{1975}).

\bibitem[{\citenamefont{H{\"u}bel}(2005)}]{Hubel2005Prog.Part.Nucl.Phys.1}
\bibinfo{author}{\bibfnamefont{H.}~\bibnamefont{H{\"u}bel}},
  \bibinfo{journal}{Prog. Part. Nucl. Phys.} \textbf{\bibinfo{volume}{54}},
  \bibinfo{pages}{1} (\bibinfo{year}{2005}).

\bibitem[{\citenamefont{Frauendorf}(2001)}]{Frauendorf2001Rev.Mod.Phys.463}
\bibinfo{author}{\bibfnamefont{S.}~\bibnamefont{Frauendorf}},
  \bibinfo{journal}{Rev. Mod. Phys.} \textbf{\bibinfo{volume}{73}},
  \bibinfo{pages}{463} (\bibinfo{year}{2001}).

\bibitem[{\citenamefont{Clark and
  Macchiavelli}(2000)}]{Clark2000Annu.Rev.Nucl.Part.Sci.1}
\bibinfo{author}{\bibfnamefont{R.~M.} \bibnamefont{Clark}} \bibnamefont{and}
  \bibinfo{author}{\bibfnamefont{A.~O.} \bibnamefont{Macchiavelli}},
  \bibinfo{journal}{Annu. Rev. Nucl. Part. Sci.} \textbf{\bibinfo{volume}{50}},
  \bibinfo{pages}{1} (\bibinfo{year}{2000}).

\bibitem[{\citenamefont{Frauendorf}(1993)}]{Frauendorf1993Nucl.Phys.A259}
\bibinfo{author}{\bibfnamefont{S.}~\bibnamefont{Frauendorf}},
  \bibinfo{journal}{Nucl. Phys. A} \textbf{\bibinfo{volume}{557}},
  \bibinfo{pages}{259c} (\bibinfo{year}{1993}).

\bibitem[{\citenamefont{Clark et~al.}(1997)}]{Clark1997Phys.Rev.Lett.1868}
\bibinfo{author}{\bibfnamefont{R.~M.} \bibnamefont{Clark}}
  \bibnamefont{et~al.}, \bibinfo{journal}{Phys. Rev. Lett.}
  \textbf{\bibinfo{volume}{78}}, \bibinfo{pages}{1868} (\bibinfo{year}{1997}).

\bibitem[{\citenamefont{Frauendorf et~al.}(1994)\citenamefont{Frauendorf, Meng,
  and Reif}}]{Frauendorf199452}
\bibinfo{author}{\bibfnamefont{S.}~\bibnamefont{Frauendorf}},
  \bibinfo{author}{\bibfnamefont{J.}~\bibnamefont{Meng}}, \bibnamefont{and}
  \bibinfo{author}{\bibfnamefont{J.}~\bibnamefont{Reif}}, in
  \emph{\bibinfo{booktitle}{Proceedings of the Conference on Physics From Large
  $\gamma$-Ray Detector Arrays}} (\bibinfo{publisher}{Berkeley},
  \bibinfo{year}{1994}), vol. \bibinfo{volume}{II of Report No. LBL35687},
  p.~\bibinfo{pages}{52}.

\bibitem[{\citenamefont{Simons et~al.}(2003)}]{Simons2003Phys.Rev.Lett.162501}
\bibinfo{author}{\bibfnamefont{A.~J.} \bibnamefont{Simons}}
  \bibnamefont{et~al.}, \bibinfo{journal}{Phys. Rev. Lett.}
  \textbf{\bibinfo{volume}{91}}, \bibinfo{pages}{162501}
  (\bibinfo{year}{2003}).

\bibitem[{\citenamefont{Simons et~al.}(2005)}]{Simons2005Phys.Rev.C24318}
\bibinfo{author}{\bibfnamefont{A.~J.} \bibnamefont{Simons}}
  \bibnamefont{et~al.}, \bibinfo{journal}{Phys. Rev. C}
  \textbf{\bibinfo{volume}{72}}, \bibinfo{pages}{024318}
  (\bibinfo{year}{2005}).

\bibitem[{\citenamefont{Datta et~al.}(2005)}]{Datta2005Phys.Rev.C41305}
\bibinfo{author}{\bibfnamefont{P.}~\bibnamefont{Datta}} \bibnamefont{et~al.},
  \bibinfo{journal}{Phys. Rev. C} \textbf{\bibinfo{volume}{71}},
  \bibinfo{pages}{041305} (\bibinfo{year}{2005}).

\bibitem[{\citenamefont{Choudhury et~al.}(2010)}]{Choudhury2010Phys.Rev.C61308}
\bibinfo{author}{\bibfnamefont{D.}~\bibnamefont{Choudhury}}
  \bibnamefont{et~al.}, \bibinfo{journal}{Phys. Rev. C}
  \textbf{\bibinfo{volume}{82}}, \bibinfo{pages}{061308}
  (\bibinfo{year}{2010}).

\bibitem[{\citenamefont{Chiara et~al.}(2000)}]{Chiara2000Phys.Rev.C34318}
\bibinfo{author}{\bibfnamefont{C.~J.} \bibnamefont{Chiara}}
  \bibnamefont{et~al.}, \bibinfo{journal}{Phys. Rev. C}
  \textbf{\bibinfo{volume}{61}}, \bibinfo{pages}{034318}
  (\bibinfo{year}{2000}).

\bibitem[{\citenamefont{Zhu et~al.}(2001)}]{Zhu2001Phys.Rev.C41302}
\bibinfo{author}{\bibfnamefont{S.}~\bibnamefont{Zhu}} \bibnamefont{et~al.},
  \bibinfo{journal}{Phys. Rev. C} \textbf{\bibinfo{volume}{64}},
  \bibinfo{pages}{041302} (\bibinfo{year}{2001}).

\bibitem[{\citenamefont{G. A. Lalazissis, P. Ring, and D. Vretenar}(2001)}]{LNP.641}
\bibinfo{author}{\bibfnamefont{G. A.}~\bibnamefont{Lalazissis,}}
  \bibinfo{author}{\bibfnamefont{P.} \bibnamefont{Ring,}} \bibnamefont{and}
  \bibinfo{author}{\bibfnamefont{D.} \bibnamefont{Vretenar}} \bibnamefont{Eds.}
  \emph{\bibinfo{title}{Extended Density Functionals in Nuclear Structure Physics,}}
  \bibinfo{title}{Lecture Notes in Physics}, vol.~\bibinfo{volume}{641}
  (\bibinfo{publisher}{Springer, Heidelberg}, \bibinfo{year}{2004}).


\bibitem[{\citenamefont{Cohen et~al.}(1992)}]{CFG.92}
\bibinfo{author}{\bibfnamefont{T. D.}~\bibnamefont{Cohen,}}
  \bibnamefont{et~al.}
  \bibinfo{journal}{Phys. Rev. C} \textbf{\bibinfo{volume}{45}},
  \bibinfo{pages}{1881} (\bibinfo{year}{1992}).

\bibitem[{\citenamefont{Peng et~al.}(2008)}]{Peng2008Phys.Rev.C24313}
\bibinfo{author}{\bibfnamefont{J.}~\bibnamefont{Peng}} \bibnamefont{et~al.},
  \bibinfo{journal}{Phys. Rev. C} \textbf{\bibinfo{volume}{78}},
  \bibinfo{pages}{024313} (\bibinfo{year}{2008}).

\bibitem[{\citenamefont{Madokoro et~al.}(2000)}]{Madokoro2000Phys.Rev.C61301}
\bibinfo{author}{\bibfnamefont{H.}~\bibnamefont{Madokoro}}
  \bibnamefont{et~al.}, \bibinfo{journal}{Phys. Rev. C}
  \textbf{\bibinfo{volume}{62}}, \bibinfo{pages}{061301}
  (\bibinfo{year}{2000}).

\bibitem[{\citenamefont{Zhao et~al.}(2011)}]{Zhao2011Phys.Lett.B181}
\bibinfo{author}{\bibfnamefont{P.~W.} \bibnamefont{Zhao}} \bibnamefont{et~al.},
  \bibinfo{journal}{Phys. Lett. B} \textbf{\bibinfo{volume}{699}},
  \bibinfo{pages}{181} (\bibinfo{year}{2011}).

\bibitem[{\citenamefont{Olbratowski
  et~al.}(2004)}]{Olbratowski2004Phys.Rev.Lett.52501}
\bibinfo{author}{\bibfnamefont{P.}~\bibnamefont{Olbratowski}}
  \bibnamefont{et~al.}, \bibinfo{journal}{Phys. Rev. Lett.}
  \textbf{\bibinfo{volume}{93}}, \bibinfo{pages}{052501}
  (\bibinfo{year}{2004}).

\bibitem[{\citenamefont{Zhao et~al.}(2010)}]{Zhao2010Phys.Rev.C54319}
\bibinfo{author}{\bibfnamefont{P.~W.} \bibnamefont{Zhao}} \bibnamefont{et~al.},
  \bibinfo{journal}{Phys. Rev. C} \textbf{\bibinfo{volume}{82}},
  \bibinfo{pages}{054319} (\bibinfo{year}{2010}).

\bibitem[{\citenamefont{Koepf and Ring}(1989)}]{Koepf1989Nucl.Phys.A493}
\bibinfo{author}{\bibfnamefont{W.}~\bibnamefont{Koepf}} \bibnamefont{and}
  \bibinfo{author}{\bibfnamefont{P.}~\bibnamefont{Ring}},
  \bibinfo{journal}{Nucl. Phys.} \textbf{\bibinfo{volume}{A493}},
  \bibinfo{pages}{61} (\bibinfo{year}{1989}).

\end{thebibliography}
\end{document}